\documentclass[11pt,a4paper]{article}
\usepackage{jheppub}

\begin{document}

\title{b DECAYS: A FACTORY FOR HIDDEN \\ CHARM MULTIQUARK STATES}
\author[]{Franco Buccella}
\affiliation[]{INFN, Sezione di Napoli,  via Cintia, 80126 Napoli, Italy}
\emailAdd{franco.buccella@na.infn.it}

%\date{February 27, 2000}

\abstract{ We assume that the two hidden charm pentaquark
discovered at $LHCb$ are built by the three light quarks 
and the $c \bar{c}$ pair, both colour octets. in relative 
P-wave for the $\frac{5}{2}^+$ state and by the eigenvectors 
of the chromomagnetic interaction for the five constituents 
in S-wave for the four $\frac{3}{2}^-$ state with masses
$4360$, $4410$, $4490$ and $4560$ MeV . The "open channel" 
$p $ $J/\psi$ has large components along the
two first states and so appear as the $4380$ resonance
and the $\Lambda_c$ $ \bar{D}^{0*}$ and $\Sigma_c$ $\bar{D}^*$
are "open channels" for the lower and higher resonances, 
respectively.
The expectation for their strange isoscalar partners is to have 
a mass larger by an amount of the order or smaller than the 
$M_{\Lambda} - M_N$ mass difference.
We account for the small width of the $\frac{5}{2}^+$, since its
decay needs the exchangeof a gluon and stress that the decays of 
particles with "beauty" provide the best experimental framework 
for the discovery of hidden charm multiquarks.
The relevance of the production mechanics shows the reason why 
only some of the states with non minimal number of constituents 
have been found.}

%\pacs{21.10-k, 21.10.Pc, 03.65.Pm}

\maketitle
%\tableofcontents
%%%%%%%%%%%%%%%%
\section{Introduction}

The discovery of two pentaquarks with hidden charm in the decay
\cite{LHC_b}:

\begin{equation}
 \Lambda_b \rightarrow p + J/\psi + K^-       
\label{equation1}
\end{equation}

confirms the attitude of particles with beauty to give rise to multiquark
states with hidden charm previously shown by the discovery of the 
$(3872, 1^+)$ resonance decaying into $J/\psi + \rho^0$ (or $\omega$) 
produced together with a caon in the decay of $B_q$'s \cite{3872}.
The study of states with non-minimal number of constituents started
about forthy years ago as well as for $2q$ $2 \bar{q}$ states 
\cite{J1} \cite{CH} as for $6q$ states \cite{J2} \cite{MS} and, last
but not least, for $4q$ $\bar{q}$ states \cite{HS} . Their existence
has received positive evidence by the analysis, which has confirmed the
existence of the $\Theta^+, Y =2$ baryon resonance \cite{As} and by 
the large cross-section at high momentum transfer of the $(3872, 1^+)$
resonance \cite{CMS}, which shows that it is a compact object \cite{EGM} 
and not a molecule, which should behave as the deuteron, which is very 
rarely produced at high momemta \cite{Alice} (I thank Prof. Antonello 
Polosa to bring this fact to my attention). This seems to 
confirm the $2q$ $2 \bar{q}$ configuration for the $X(3872)$, 
as explicated in\cite{HRS} (see also \cite{BHRS}).
The discovery of the two hidden charm pentaquarks gave, if
necessary, a new encouragement to deepen the theoretical study
of these particles. Here we assume that their spectrum may be 
described in terms of the chromomagnetic interaction (CMI), 
which has been successful to describe the mass differences 
of the states of the $56$ of $SU(6)$ flavor spin (the octet 
$\frac{1}{2}^+$ and the decuplet $\frac{3}{2}^+$) \cite{DGG}.
Let us begin by the mechanism of their formation in  the Cabibbo 
allowed process for the decaying $\Lambda_b$ with amplitude
proportional to $V^*_{cb} V_{cs}$:

\begin{equation}
b \rightarrow c + s + \bar{c}            
\label{equation2}
\end{equation}

To give rise to the seven constituents a gluon should be
emitted and converted into a  $u \bar{u}$ pair. To produce
the final $K^-$ the $\bar{u}$ should combine with the strange
quark produced in the decay, while the $u$ may form with
the spectator scalar and isoscalar diquark in the $\Lambda_b$ by 
color conservation an octet of color anf flavor with spin 
parity $\frac{1}{2}^+$, which may combine with the pair $c \bar{c}$, 
a color octet with spin one, produced in the decay to form the 
pentaquarks.
If the five constituents join in relative S-wave, they may
give rise to the $\frac{3}{2}^-$ hidden charm pentaquark, 
while if the two octets are in a P-wave they may give rise
to the $\frac{5}{2}^+$ . In the first case, when they join, 
they give rise to a combination of eigenstates of the QCD 
hamiltonian. 
In both cases isospin conservation requires that the three 
light quarks have $I = \frac{1}{2}$.
Therefore Pauli principle demands that, if they are in S-wave with a
symmetric wave function, they transform as the $70$ representation of 
$SU(6)_{cs}$ (for three objects a mixed symmetry may give rise to an
antisymmetric one only by multiplying it by another mixed symmetry,
since by multiplying it for a totally symmetric or antisymmetric one
gets a mixed symmetry \cite{HS1} \cite{CHS} \cite{BS}), while the $c 
\bar{c}$ pair transforms as the $ 35 + 1 $ representation of $SU(6)_{cs}$ .
The mechanism of the decay of the $(\frac{5}{2})^+$ pentaquark
is similar to the one, which may describe the decay of the
$1^+$ tetraquark at $3872$ into $J/psi + \rho^0$ (or $\omega$):
a gluon exchange, which turns the two color octets into singlets.
Instead their formation in the decay of $B_q$ and $\Lambda_b$
is different. In fact in the first case the strange quark
produced in the $b$ decay forms a caon together with the
spectator antiquark, while the $ q \bar{q}$ pair produced by
the gluon forms together with the $c \bar{c}$ the $1^+(3872)$
tetraquark. The analogous process for the $\Lambda_b$ decay,
with the strange quark forming with the spectator diquark a
$\Lambda$ would give rise to the decay 
$\Lambda_b \rightarrow \Lambda + 1^+(3872)$
which might be looked for in final states $p + \pi^- (\Lambda)$,
$\mu^+ \mu^- (J/\psi)$ and $\pi^+ \pi^- (\rho^0)$.
H\"{o}gaasen and Sorba \cite{HS1} studied all the possibilities 
with three constituents in P-wave with respect to the other two 
and came to the conclusion that the most interesting case is with 
two color octets of the thre light quarks and the $c$ $\bar{c}$ pair with 
the caveat that each octet might be turned into an ordinary hadron by 
absorbing a gluon before combining to form the hidden charm pentaquark. 
One should keep however into account the fact that in the decay 
$\Lambda_b \rightarrow p + J/\psi + K^-$
a gluon should be emitted and turned into a $u$ $\bar{u}$ pair to
give rise to seven final constituents and therefore the presence
of another gluon requires a higher order in QCD. As we shall stress 
in the following the "beautiful" particles due to their relative 
long lifetime decay at a distance from the interaction 
point sufficient to avoid the presence of the gluons emitted there.
In the next section we shall show the role of the chromomagnetic 
interaction (the fine structure term) to describe the spectrum
of the ordinary hadrons and of the two lowest scalar nonets and 
the mass of the doubly charmed baryon recently discovered at
$LHCb$ \cite{LHC_b1}.
In the third section we will compute the masses of the
two hidden charm pentaquarks.
Our description will account for their different widths. 
In the fourth  section we shall give a reason for which
only some of the multiquark states have been detected. 
Finally we shall give our conclusion.  
In the Appendix we write some CG of $SU(6)$ and some identities,
which are useful to compute the chromomagnetic contributions.

\section{The spectrum of the lower negative and positive
parity mesons described by the chromomagnetic interaction.}

After the proposal of QCD as the theory of strong interactions
\cite{FGL} De Rujula, Georgi and Glashow \cite{DGG} realized 
that the fine structure (the chromomagnetic interaction) accounts 
for the mass differences between $\Delta$ and the nucleon and
between $\Sigma$ and $\Lambda$. In the same framework there
is the successful prediction:

\begin{equation}
M(\Xi^*) - M(\Xi) = M(Y^*) - M(\Sigma)
\label{equation3} 
\end{equation}

which was previously obtained by assuming the same coefficients
for the terms trasforming as an octet for the decimet and the octet
baryons.
By applying the same approach to the charmed baryons 
$\Sigma_c$ and $\Lambda_c$ they predict a mass difference high
enough to allow the strong decay 
$\Sigma^+_c \rightarrow \Lambda_c + \pi^+$  in agreement with the
discovery of both particles in a neutrino experiment \cite{CaC}
(I am grateful to Professor Alvaro De Rujula to bring 
\cite{CaC} to my attention) .
Indeed the masses of these two particles are reproduced with a
girochromo-magnetic factor $k_c = 0.24$ $k_u$ and with an effective 
mass for the charmed quark $1715$ MeV. The $\Sigma_b$ and
$\Lambda._b$ particles have a mass difference even larger, as
expected . The mass of $3621.40$ MeV of the $\Xi^{++}_{cc}$ 
recently found by $LHCb$ \cite{LHC_b1} implies an effective 
mass of the constituent charmed quarks of $1665 MeV$ smaller 
than the one found for the charmed mesons and  $\Lambda_c$. 
As long as for mesons ($\pi$, $K$ $\rho$, $K^*$),one obtains 
their masses with a larger effective CMI and smaller effective 
masses for the light and the strange quarks. Both these
properties can be understood by the more intense chromo-electric
actraction between a quark and an antiquark, which form a color 
singlet with respect to two quarks, which combine in a $\bar{3}$ 
of SU(3) color . Indeed the stronger actraction implies a smaller
constituent mass and a larger contact interaction.
In fact for the charmed mesons $D$ and $D^*$ a slightly smaller mass,
$1615$ MeV, and larger $k_c = 0.26$ $k_u$ are needed with respect to 
the charmed baryons .
Also the values found for the $c$ $\bar{c}$ states, 1535 MeV for the 
mass of the charmed quark and $k^2 = 0.186$ for the square of the 
giro chromomagnetic factor can be understand as a consequence of the 
smaller distance between the constituents.
For the two nonets of scalar tetraquarks, where the states built
with the light constituents are the $f^0(600)$ and $f^0(1370)$,
their masses are reproduced with an effective chromomagnetic
interaction as for the baryons and with a larger constituent mass.
Interestingly enough one explains why the lowest one, which decays
into two pions, has a very large width, while the other one decays 
mainly into four pions \cite{Gaspero}. 
In fact the SU(6) color spin Casimir, which gives the most important 
contribution to the masses, implies that the ligther state is almost a 
SU(6) color spin singlet with an "open channel" \cite{J} into two pions, 
which are also color singlets, while the heavier one transforms mainly 
as a 405 and therefore has an open channel into a pair of $\rho$ mesons, 
which transform as a 35 of SU(6) color spin \cite{B} \cite{BHRS} .
We may be confident that also the pentaquark states are
eigenvectors of the chromomagnetic interaction.
A general analysis of the spectrum of negative and positive pentaquarks
built with the three lightest quarks can be found in \cite{ABFRT} and
the study of $3q$ $3 \bar{q}$ exaquarks in \cite{ABT}. 
   
\section{Formation, masses and decays of the hidden charm
pentaquarks}

In the Cabibbo allowed process for the decaying $\Lambda_b$ 
described in (2) the produced $c$, if it does not recombine 
with the spectator diquark $u d$ to give a $\Lambda_c$, may 
form a color octet with spin $1$ with the $\bar{c}$. If a 
gluon is produced and which gives rise to a color octet $u \bar{u}$ 
pair, the $\bar{u}$ may combine with the s produced in the decay to 
form a negative kaon and the $u$ with the spectator diquark in the 
$\Lambda_b$ may form a color octet with spin $\frac{1}{2}$ and the 
same flavor of the proton. 
The two color octets may give rise to one or the other of the two 
resonances, depending on their relative orbital momentum, the one 
with negative parity for the S-wave, the one with positive parity 
for the P-wave. We assume that the mass $= 4450$ MeV of the 
$(\frac{5}{2})^+$ pentaquark is given by the sum of the constituent 
masses and of the contributions of  the chromomagnetic interaction (CMI), 
which for the usual baryons reproduces the $M_{\Delta}-M_N$ and the 
$M_{\Sigma}-M_{\Lambda}$ mass differences  \cite{DGG} and for the 
charmed mesons $M_{ J/ \psi}-M_{\eta_c}$ , of the rotational 
energy and of the spin orbit $\vec{L} \times \vec{S}$ terms. 
Within the semplifying assumption that the contribution of the
chromoelectric interaction of four $3$ and a $\bar{3}$ does not 
depend on the way they form a color singlet, we can relate the mass
of the two octets to two combinations of the mass of the $\Delta$
and the $N$ for the three light quarks and of $J/\psi$ and $\eta_c$
for the $c \bar{c}$ pair .

Starting by the formulas:

\begin{equation}
- \frac{m_{\Delta}- m_N}{4} 
[ C(3q)_6 - \frac{1}{2} C(3q)_3 - \frac{1}{3} C(3q)_2 - 6 ]
\label{equation4}
\end{equation}

($C_n$ are the quadratic Casimir of $SU(6)_{cs}$, $SU(3)_{c}$ and 
$SU(2)_{s}$,respectively) for the three quarks and:

\begin{equation}
\frac{3}{16}
(M_{J/\psi}- M{\eta_c})
[ C(c \bar{c})_6 - \frac{1}{2} C(c \bar{c})_3 - \frac{1}{3} C(c \bar{c})_2 
- 4 ]
\label{equation5}
\end{equation}

for the $c$ $\bar{c}$ pair.

By adding the contribution of the constituent masses:

\begin{equation}
\frac{M_N + M_{\Delta}}{2} + \frac{3 M_{(J/\psi} + M_{\eta_c)}}{4} = 
(1085.5 + 3068) MeV = 4153.5 MeV
\label{equation6}
\end{equation}

the sum of the masses of the two octets, which build the $(\frac{5}{2})^+$
is given by: 

\begin{equation}
\frac{3 M_{\Delta} + 5 M_N}{8} + \frac{23 M_{J/\psi} + 9 M_{\eta_c}}{32} 
= 4127 MeV 
\label{equation7}
\end{equation}

The spin-orbit terms, which are both expected to be positive, and the 
rotational energy may give the remaining contribution.
The narrow width of the $4450$ MeV, $\frac{5}{2}^+$ should be explained 
by the fact that the decay into $ p + J/\psi $ needs the exchange of one 
gluon as it was the case for the decay of the $1^- (3872)$ into $J/\psi + 
\rho^0$ (or $\omega$), if one identifies it as the state built with the
light $(q \bar{q})$ and the charmed $(c \bar{c})$ pairs transforming 
as the $(8, 3)$ representation of $SU(3)\times SU(2)$ color spin 
\cite{HRS}. 
One has with all the constituents in S-wave four states with 
$S = \frac{3}{2}$ , which can be obtained by the products 
$\frac{3}{2} \times 1$, $\frac{3}{2} \times 0$ and 
$\frac{1}{2} \times 1$ . 
Let us remember that the $70$ contains both $\frac{3}{2}$  and 
$\frac{1}{2}$ spin color octets and a spin $\frac{1}{2}$ singlet, 
while the $35$ contains both $1$ and $0$ spin color octets and a spin $1$ 
color singlet. 
So we have the following possibilities for the color-spin transformation 
properties of the three light quarks and the $ c \bar{c} $ pair:

\begin{equation}
(8, \frac{3}{2}) \times (8, 1) \textrm{ combined into a } (1, \frac{3}{2})  
\label{equation8}
\end{equation}

\begin{equation}
(8, \frac{3}{2}) \times (8, 0) \textrm{ combined into a } (1, \frac{3}{2}) 
\label{equation9}
\end{equation}

\begin{equation}
(8, \frac{1}{2}) \times (8, 1) \textrm{ combined into a } (1, \frac{3}{2})  
\label{equation10}
\end{equation}

\begin{equation}
(1, \frac{1}{2}) \times (1, 1) \textrm{ combined into a } (1, \frac{3}{2}) 
\label{equation11}
\end{equation}

One has also to consider the chromomagnetic interaction 
between the charmed and the light quarks and their different chromomagnetic 
factors \cite{HS2} \cite{CHS} .
The total contribution of CMI, M, is given by:
\begin{equation}
M = M(70)+ M(70 \times 6) + M(70 \times \bar{6}) + 
M(6 \times \bar{6})
\label{equation12}
\end{equation}

The sum of the contributions of the first and the fourth terms 
to the states defined in eqs.(8...11) is given by:

\begin{equation}
\frac{m_{\Delta}- m_N}{8} -  \frac{m_{J/\psi}- m_{\eta_c}}{32}
\label{equation13}
\end{equation}

for (8)  

\begin{equation}
 \frac{m_{\Delta}- m_N}{8} +  \frac{3(m_{J/\psi}- m_{\eta_c}}{32}
\label{equation14}
\end{equation}

for (9)

\begin{equation}
- \frac{M_{\Delta}- M_N}{4} - \frac{M_{J/\psi}- M_{\eta_c}}{32}
\label{equation15}
\end{equation}

for (10)

while for the "open channel" (11) is:

\begin{equation}
\frac{M_N - M_{\Delta}}{2} + \frac{M_{J/\psi} - M_{\eta_c}}{4} 
\label{equation16}
\end{equation}

The second and the third term, related to the chromomagnetic
interaction of the light quarks with $c$ and $\bar{c}$, are 
proportional to $k_1 = 0.24$ and $k_2 = 0.26$, respectively . 
To evaluate them one should consider the tensor products:

\begin{equation}
70 \times 6 = 210 + 105 +105'
\label{equation17}
\end{equation}

\begin{equation}
70 \times \bar{6} = 384 + 21 +15
\label{equation18}
\end{equation}

and the fact that the $(3,5)$ of $SU(6)_{cs}$ is contained in the
$105'$, while the three $(3,3)$ in the three representation of the
first product, and that one of the $(\bar{3},3)$ is contained in the
$15$ and the $(\bar{3},5)$ and the other two $(\bar{3},3)$'s in the 384
for the second product .
In conclusion the terms proportional to:

$(M_{\Delta} - M_{N})$ 

and

$(M_{\rho} - M_{\pi}) = \frac{1}{k^2}  (M_{J/\psi} - M_{\eta_c})$

are the matrices:

\begin{table}[h]

\begin{center}

\begin{tabular}{|c|c|c|c|}

\hline
 
$\frac{1 + 3k_1}{8}$  &  0  &  0  &  0  \\ 

\hline

 0  &  $\frac{1 - 3k_1}{8}$  &  $\frac{k_1}{3}$  &  $\frac{k_1}{6}$ \\ 

\hline

 0  &  $\frac{k_1}{3}$  &  $ - \frac{1 + 3k_1}{8}$  &  $\frac{k_1}{6}$ \\ 

\hline
 0  &  $\frac{k_1}{6}$ & $\frac{k_1}{6}$ & $\frac{k_1 - 1}{2}$ \\
 
\hline

\end{tabular}

\end{center}

\caption{\footnotesize{}}

\end{table}

\begin{table}[h]

\begin{center}

\begin{tabular}{|c|c|c|c|}

\hline
 
 $\frac{3 k_2 (-9 + k_2}{64}$ &  - $\frac{\sqrt{15}(3+k_2)}{64}$  & 
- $\frac{\sqrt{15}k_2}{8}$  & - $\frac{\sqrt{15}k_2}{16}$  \\

\hline

- $\frac{\sqrt{15}(3+k_2)}{64}$  & - $\frac{(15 + k_2)k_2 }{64}$  & 
- $\frac{\sqrt{15}k_2}{8}$  & - $\frac{\sqrt{15}k_2}{16}$  \\

\hline

- $\frac{\sqrt{15}k_2}{8}$  & - $\frac{k_2}{8}$  &  
- $\frac{(3 - k_2)k_2 }{32}$  & - $\frac{k_2}{8}$  \\

\hline

- $\frac{\sqrt{15}k_2}{16}$ & - $\frac{k_2}{16}$  &
- $\frac{k_2}{8}$  &  $\frac{(k_2)^2}{4}$ \\

\hline

\end{tabular}

\end{center}

\caption{\footnotesize{}}

\end{table}

respectively .

We get the following M matrix in MeV for the total contribution of
the chromomagnetic interaction in the base of the states:
\begin{eqnarray}
| 1 > &=& |  70 \times 6,  (8,4) \times (3,2) \rightarrow (3,5) > \\ \nonumber
| 2 > &=& |  70 \times 6,  (8,4) \times (3,2) \rightarrow (3,3) > \\ \nonumber
| 3 > &=& |  70 \times 6,  (8,2) \times (3,2) \rightarrow (3,3) > \\ \nonumber
| 4 > &=& |  70 \times 6,  (1,2) \times (3,2) \rightarrow (3,3) >
\end{eqnarray}

\begin{table}[h]

\begin{center}

\begin{tabular}{|c|c|c|c|}

\hline
 
 10.5  &  - 33  &  - 71.5  &  - 35.05  \\ 

\hline

- 33  &  28.75  &   5.3   & 2.65 \\ 

\hline
- 71.1  &  5.3  & - 53.2   & 6.5 \\ 

\hline
- 35.05  &  2.65  &  6.5   & - 86.2 \\ 

\hline

\end{tabular}

\end{center}

\caption{\footnotesize{}}

\end{table}

which has the eigenvalues: - 120, - 71, 11 and 80
in correspondence to the eigenvectors:

(.057, .08, .59,  .624 )

(.225, .063, .604, - .762 )

(.39, .847, - .35, - .094 )

(.736, -.522, .- .041, -.15 )

respectively. 
The "open channel" $p + J/\psi$ has negligible components along
the two eigenvctors corresponding to the two higher eigenvalues
and sostantial ones along the two lower ones. This complies well
with  the mass of the $(\frac{3}{2})^-$ state if we take the 
constituent masses of the light quarks from the lowest baryons
and of $c$ and $\bar{c}$ from the $\Lambda_c$ and from the lowest 
charmed mesons, respectively: 

\begin{equation}
\frac{M_{\Delta} - M_N}{2} + M_{\Lambda_c} + 
\frac{3 M_{D^*} + M_{D}}{4} = 4480 MeV
\label{equation19}
\end{equation}

which implies for the two lightest $(\frac{3}{2})^-$ states
a mass of 4360 and 4410~MeV. Indeed, by taking the masses
of charmed constituents from charmonium would lead to
smaller constituent masses, but the presence of the three
light quarks favors to consider charmed baryon and mesons
and the tendency of larger constituent masses with the
increasing number of constituents in relative S-wave
may give rise to a global constituent mass so well
reproducing the experimental value. Indeed the $Q^2$ dependence of 
the strong coupling constant, decreasing with the scale, might be an 
explanation for the different values of the constituent masses for the 
ordinary mesons and baryons as well as for the different value at the 
scale of the negative parity states built with all the constituents 
( $3 q $, $c$ and $\bar{c}$) in S-wave.
Indeed for the lowest scalar tetraquarks built with light 
constituents the effective masses of the constituents is
about $400$~MeV, larger than in the case of ordinary baryons.
The value found has the important consequence to predict two higher
$(\frac{3}{2})^-$ states at $4490$ and $4560$ MeV.
By considering $qqc$-$\bar{c}q$ combinations the "open channels"
$\Lambda_c \bar{D^{*0}}$  and the $I = \frac{1}{2}$ combination
$\frac{1}{\sqrt{3}}( \sqrt{2}$ $\Sigma^{++}_c \bar{D^{*-}}
- \Sigma^+_c \bar{D^{*0}}$)
have different components along the CMI eigenvectors.
While $\Lambda_c \bar{D^{*0}}$ with total spin $\frac{1}{2}$ for the
light quarks is a combination of the two last vectors and therefore has 
substantial components along the two lower mass eigenstates, 
$\frac{1}{\sqrt{3}}( \sqrt{2} \Sigma^{++}_c \bar{D^{*-}})
- \Sigma^+_c \bar{D^{*0}}$
has components mainly along the two 
states with spin $S(uud) = \frac{3}{2}$ for the light quarks, as it 
can be seen from the identity for the states with $S = \frac{3}{2}$:
\begin{eqnarray}
| S(uu) = 1, S(uuc) = \frac{1}{2}, S(c \bar{c}) = 1  > &=& 
\frac{1}{3} [\sqrt{5} | S(uud) = \frac{3}{2}, |S(c \bar{c}) = 1 > - 
\sqrt{3} | S(uud) = \frac{3}{2} S(c \bar{c}) = 0 > \nonumber \\ 
&&+ | S(uud) = \frac{1}{2}, S(c \bar{c}) = 1 >]
\end{eqnarray}
The relationship between the $ (8 \times 8)_1 $ and $ 1 \times 1$ for
$(uud)$ and $c \bar{c}$ and $(uuc)$ $d \bar{c}$ is supplied by the 
well known $SU(3)$ identities:
\begin{eqnarray}
\delta^{\beta}_{\alpha}  \delta^{\epsilon}_{\gamma} &=& 
\frac{1}{3} \delta^{\epsilon}_{\alpha} \delta^{\beta}_{\gamma} 
+ \frac{1}{2} (\lambda_a)^{\epsilon}_{\alpha} (\lambda_a)^{\beta}_{\gamma} \\
(\lambda_a)^{\beta}_{\alpha} (\lambda_a)^{\epsilon}_{\gamma} &=& 
\frac{16}{3} \delta^{\epsilon}_{\alpha} \delta^{\beta}_{\gamma} 
- \frac{1}{3} (\lambda_a)^{\epsilon}_{\alpha} (\lambda_a)^{\beta}_{\gamma}
\end{eqnarray}
The fact that the chromomagnetic interaction for the light quarks
(the ones with the higher girochromomagnetic factor ) gives a positive
contribution for the state $\Sigma_c$ $\bar{D}^*$ and negative for
$\Lambda_c$ $\bar{D}^{0*}$ (in analogy with the large difference
$M_{\Sigma_c}$ - $M_{\Lambda_c}$ \cite{DGG} \cite{CaC}) leads us to guess
that the $\Sigma_c$ $\bar{D}^*$ and $\Lambda_c$ $\bar{D}^{0*}$
"open channels have large components along the $4560$ and $4360$ MeV
resonances, respectively . However, according to the formation mechanism
starting from the third state $ (8, \frac{1}{2}) \times (8, 1) $, which 
has
a negligible component along the higher eigenstate, the $\Sigma_c 
\bar{D}^*$ decay may be more easily seen for the $4490$ resonance. 
For the decay of the $\Sigma^{++}_c$ we may have the same sequence:

$\Sigma^{++}_c \rightarrow \Lambda_c + \pi^+$,

$\Lambda_c \rightarrow p + K^- \pi^+$

which lead to the discovery of $\Sigma^{++}_c$ in a
neutrino experiment \cite{CaC} . \\
If it is the strange quark produced in the weak decay to form a 
strange color octet together with the scalar and isoscalar
spectator in $\Lambda_b$, similar to the description of the 
formation of the $3872$ $1^+$ in B decays \cite{A}, we can give
a qualitative description of the spectrum of the strange isoscalar
pentaquark with hidden charm. As long as for the $\frac{5}{2}^+$
the CMI interaction for the color octet $\frac{1}{2}^+$ is the same
as for the $udu$ case and we should simply add the $M_{\Lambda} -
M_N$ mass difference. We expect however smaller positive
contributions from the rotational energy and the spin-orbit
terms and therefore predict the upper limit $4625$ MeV.
For the $\frac{3}{2}^-$ states the contribution of the light
quarks, the ones with the higher girochromomagnetic factor,
to the CMI is the same and so we expect approximately for the 
four $(\frac{3}{2})^-$ states the following values for the masses:
$4535$, $4585$ , $4665$ and $4735$ MeV, respectively. \\ 
In general it is not easy to produce hadrons with non minimal 
number of constituents, since the $q$ and $\bar{q}$ produced
by the gluons tend fastly to combine into color singlets and 
the easiest way is to form ordinary hadrons. In Cabibbo allowed B 
decays the creation of a $c \bar{c}$ color octet pair, which exerts 
an actraction on another octet built with a $q \bar{q}$ pair or three 
light quarks, can give rise to hadron states with hidden charm.  \\
In conclusion the interpretation of the two pentaquark resonances
with hidden charm discovered at $LHCb$ \cite{LHC_b} as built with
a $c \bar{c}$ and three light quark color octets in P-wave for the
$(\frac{5}{2})^+$ and with the five constituents in S-wave for the
$(\frac{3}{2})^-$ accounts for their different widths.
An important consequence of this description is the prediction of
two $(\frac{3}{2})^-$ resonance at a mass of $4360$ and $4560$ MeV,
with large components along the "open channels"
$\Lambda_c$ $\bar{D}^{+0}$ and $\Sigma_c$ $\bar{D}^*$ final states, 
respectively .
As we shall stress in the following section the "beautiful"
particle due to their relative long lifetime decay at a distance
from the interaction point sufficient to avoid the presence
of the gluons emitted there, which give rise to $q$ $\bar{q}$ 
pairs transforming as color octets with the $q$'s and the
$\bar{q}$'s, which build with the other constituents
ordinary hadrons.

\section{Formation of multiquark states}

The fact that the $3872, 1^+$, which is a compact object, since it 
is produced also at high $p_T$ at difference from the deuton
is seen only for its neutral component shows the relevance of
the formation of multiquark states. In fact the mechanism described is 
operative only for the neutral component \cite{A}, which is indeed the
only component discovered. \\
In fact as long as for the  states predicted by Jaffe \cite{J} strong 
evidence concerns only the two scalar nonets, the multiplet consisting of 
$f^0(600)$, $k(770)$ and $f'^0$ and $A^0$, degenerate as expected, at 980 
MeV and the one, where $f^0(1370)$ is the one consisting of light 
constituents. This lead the Roma group \cite{MPPR} to consider only the 
diquarks transforming as $(\bar{3}, S=0, \bar{3})$ with respect to 
$SU(3)_c \times SU(2)_s \times SU(3)_f$ and their antiparticles, which may 
give rise only to one scalar nonet. 
To account for the heavier one they advocate an istanton \cite{t'H}. As we 
have shown in the second section the masses and decays of the two states 
built with the light constituents are well described by deducing their 
spectrum with the same approach followed in \cite{DGG} for ordinary 
hadrons. In fact, when the $\bar{3}, S=0$ and $3, S=0$ join, they give 
rise to a superposition of eigenstates of the CMI, with "open channel" 
\cite{J} two pions or two $\rho$'s, respectively.
To build the $3872$ $1^+$ the Roma group considered also diquarks
transforming as a $(\bar{3}, 3)$ under $SU(3)_c x SU(2)_s$ \cite{MPPR2}. 
For these diquarks the chromoelectric force is actrative, while the 
chromomagnetic is repulsive, which makes their formation less probable.
Moreover, as well as the diquark $(\bar{3}, 1)$, they may combine with
a quark to form a baryon. The $(\bar{3},3)$ may also give rise to a
flavor decimet and therefore a lower limit to the ratio of the abundances
of $(\bar{3},3)$ and $\bar{3}, 1)$ may be given by the ratio of the 
non-diffractive productio of $\Delta$ and N .
Diquarks are considered to build tetraquarks \cite{MPPR3} as well as for
pentaquarks \cite{MPR} with a description different from the one 
presented here. The mechanism proposed here for the formation of the 
$3872, 1^+$, which accounts for the fact that only its neutral 
component has been found, is at our advice better motivated.
The tendency of the diquark $(\bar{3}, 3)$ to form a baryon with
the quark rather than combine with a $(\bar{3}, 1)$ diquark to 
build a spin 1 state or with a $(\bar{3}, 3)$ diquark
to give rise to spin and (or) isospin 2 states explains why the 
large class of states predicted by Jaffe has not yet been found .
At our advice the approach based on the extension to the 
multiquark states of the one introduced in \cite{DGG} for the
ordinary baryons, which may be successfully extended to ordinary
mesons, to find their spectrum is valid, but a production mechanism
is needed to prevent that the formation of those states is
not overwhelmed by the recombination of the q and $\bar{q}$  
produced by the gluons into ordinary hadrons . To this extent
the decays of the particles with beauty produced at Belle and
BaBaR, but also at $LHCb$ is a favourable situation, since the
"beautiful" particles decay in absence of associated production.
This is evident for the $e^+ e^-$ rings, but it happens also
for the particles produced at $LHCb$, since the long lifetime
of the b quark, which allows the hadrons with beauty to leave
the interaction point before decaying, implies that the products
of their decays are not surrounded by the $q \bar{q}$ pairs
produced in the interaction.  
As long as for the formation of the $\Xi^{++}_{cc}$ previously 
mentioned \cite{LHC_b1}, it is probably due to the union of a 
$cc$ scalar diquark with a $u$. While its decay into $\Lambda_c
+ K^- + \pi^+ \pi^+$ requires that the allowed Cabibbo decay
is accompanied by the creation of both a $u \bar{u}$ and a
$d \bar{d}$ pairs, the $\Lambda_c \rightarrow P + K^- + \pi^+$
implies the formation of a $u \bar{u}$ pair. 

\section{Conclusions.}

The approach based on the chromomagnetic interaction to find
the spectrum of the multiquark states, applied successfully
for the $3872, 1^+$ \cite{HRS} and to the lowest scalar nonets
\cite{BHRS}, is confirmed by the discovery of the two hidden
charm pentaquarks at $LHCb$, since it accounts for the different
widths of the $\frac{3}{2}^-$ and $\frac{5}{2}^+$ resonances.
A confirm of the description proposed here should be the detection
of $\Sigma_c$ $\bar{D}^*$ and $\Lambda_c$ $\bar{D}^{*0}$ final 
states and of isoscalar strange hidden charm pentaquarks with
a mass around or minor than the sum of the mass of their non strange 
partners and the mass difference $M_{\Lambda} - M_N$ .
The property of the beautiful particles  of travelling away from the 
interaction point, as a consequence of their lifetime, prevents 
the formation of hidden charm multiquarks after the Cabibbo favored 
decay with the production of a pair $c \bar{c}$  from being overwhelmed 
by the production of ordinary hadrons, which is the reason why many 
multiquark states have not been found.  \\

\section{Aknowledgement.}

I am very grateful to Prof. Mario Abud for his explanation of
the non detection of the charged partners of the 3872, which
ispired the considerations on the relevance of the formation
mechanism of multiquark states.

%\section*{references}

\section{Appendix}

The evaluation of the contributions proportional to
$k_1$, $k_2$ and $k^2$ require the knowledge of the
following CG of $SU(6)$ color spin:
\begin{eqnarray}
| 105', 3, S = 2 > &=& |70 \times 6, 8 \times 3, 
\frac{3}{2} \times \frac{1}{2} >  \nonumber \\
| 105', 3, S = 1 > &=& \frac{1}{\sqrt{6}}
\left[ |70 \times 6, 8 \times 3, \frac{3}{2} \times \frac{1}{2} > 
+ |70 \times 6, 8 \times 3, \frac{1}{2} \times \frac{1}{2} > 
+ 2 |70 \times 6, 1 \times 3, \frac{1}{2} \times \frac{1}{2} > \right] 
\nonumber \\
| 105, 3, S = 1 > &=& \frac{1}{\sqrt{3}}
\left[ |70 \times 6, 8 \times 3, \frac{3}{2} \times \frac{1}{2} > 
+ |70 \times 6, 8 \times 3, \frac{1}{2} \times \frac{1}{2} > 
- |70 \times 6, 1 \times 3, \frac{1}{2} \times \frac{1}{2} > \right] 
\nonumber \\
| 210, 3, S = 1 > &=& \frac{1}{\sqrt{2}}
\left[ |70 \times 6, 8 \times 3, \frac{3}{2} \times \frac{1}{2} > 
- |70 \times 6, 8 \times 3, \frac{1}{2} \times \frac{1}{2} > \right] 
\nonumber\\
| 384, \bar{3}, \bar{S} = 2 > &=& | 70 \times \bar{6}, 8 \times \bar{3},
\frac{3}{2} \times \frac{1}{2} >  \nonumber \\
| 384_1, \bar{3}, \bar{S} = 1 > &=& \frac{1}{\sqrt{105}} 
\left[ 5 | 70 \times \bar{6}, 8 \times \bar{3}, \frac{3}{2} 
\times \frac{1}{2} >
- 8 | 70 \times \bar{6}, 8 \times \bar{3}, \frac{1}{2} \times \frac{1}{2}  > 
- 4 | 70 \times \bar{6}, 1 \times \bar{3}, \frac{1}{2} \times \frac{1}{2}  > 
\right] \nonumber \\
| 384_2, \bar{3}, \bar{S} = 1 > &=& \frac{1}{\sqrt{5}}
\left[ | 70 \times \bar{6}, 8 \times  \bar{3}, \frac{1}{2} \times \frac{1}{2} >
- 2 | 70 \times \bar{6},  \times \bar{3}, \frac{1}{2} \times 
\frac{1}{2} > \right]   \nonumber \\
|  15, \bar{3}, \bar{S} = 1 > &=& \frac{1}{\sqrt{21}} 
\left[ 4 | 70 \times \bar{6}, 8 \times \bar{3}, \frac{3}{2} \times 
\frac{1}{2} >  
+ 2 | 70 \times \bar{6}, 8 \times \bar{3}, \frac{1}{2} \times \frac{1}{2}  > 
+ 1 | 70 \times \bar{6}, 1 \times \bar{3}, \frac{1}{2} \times \frac{1}{2}  >
\right] \nonumber
\end{eqnarray}

where $S$ is the total spin of the three light quarks and of $c$
and $\bar{S}$ is the total spin of the three light quarks and of 
$\bar{c}$. If the spin of the light quarks and the total spin are
both $\frac{3}{2}$ the following identities follow:
\begin{eqnarray}
| S = 2 > &=&  - \frac{1}{4} [ | \bar{S} = 2 > + \sqrt{15} 
| \bar{S} = 1 > ] = \frac{1}{\sqrt{8}} 
[ \sqrt{3} | S_{c \bar{c}} = 1 > +  \sqrt{5} | S_{c \bar{c}} = 0 > ] 
\nonumber \\
| S = 1 > &=& \frac{1}{4} [ | \sqrt{15} \bar{S} = 2 > 
- | \bar{S} = 1 > ] = \frac{1}{\sqrt{8}} 
[ \sqrt{5} | S_{c \bar{c}} = 1 > - \sqrt{3} | S_{c \bar{c}} = 0 > ] 
\end{eqnarray}
\end{document}